\documentclass{article}
\usepackage{arxiv}

\usepackage[utf8]{inputenc} 
\usepackage[T1]{fontenc}    
\usepackage{hyperref}       
\usepackage{url}            
\usepackage{booktabs}       
\usepackage{amsfonts}       
\usepackage{nicefrac}       
\usepackage{microtype}      
\usepackage{cleveref}       
\usepackage{lipsum}         
\usepackage{graphicx}
\usepackage{natbib}
\usepackage{doi}
\title{A $1$-D Special Relativistic \\ Shock Tube where Stellar Fluid \\ undergoes Neutrino Heating}

\date{October 27, 2023}

\newif\ifuniqueAffiliation
\uniqueAffiliationtrue

\ifuniqueAffiliation 
\author{ \href{https://orcid.org/0009-0005-4128-6994}{\includegraphics[scale=0.06]{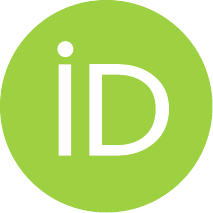}\hspace{1mm}Gregory~Mohammed}\thanks{\href{https://github.com/gregorymohammed/Gregory_Mohammed_Simulation_Code_Masters}{The simulation code in Java}} \\
	\\ Department of Physics\\
	University of British Columbia - Okanagan Campus\\
	Kelowna, BC. Canada V1V 1V7 \\
	\texttt{gregory.mohammed@ubc.ca} \\
}
\else
\usepackage{authblk}

\newbox{\orcid}\sbox{\orcid}{\includegraphics[scale=0.06]{orcid.pdf}} 
\author[1]{%
	\href{https://orcid.org/0009-0005-4128-6994}{\usebox{\orcid}\hspace{1mm}Gregory~Mohammed\thanks{\texttt{gregory.mohammed@ubc.ca}}}%
}
\affil[1]{Department of Physics, University of British Columbia - Okanagan Campus, Kelowna, BC. Canada V1V 1V7}
\fi

\renewcommand{\headeright}{}
\renewcommand{\undertitle}{}
\renewcommand{\shorttitle}{A $1$-D Special Relativistic Shock Tube where Stellar Fluid undergoes Neutrino Heating}

\hypersetup{
pdftitle={A template for the arxiv style},
pdfsubject={q-bio.NC, q-bio.QM},
pdfauthor={David S.~Hippocampus, Elias D.~Striatum},
pdfkeywords={First keyword, Second keyword, More},
}

\begin{document} 
\maketitle

\begin{abstract}
   {To investigate the physical nature of neutrino-heating on the result of a 1-dimensional core-collapse supernova.}
   {\citep*{colgate} were the first to suggest that neutrinos may play a crucial role in core collapse supernova by taking up gravitational binding energy from the core and depositing it in the rest of the star. The fluid is contained in a shock tube in a region extending from the neutrino-sphere of a star out $1000$~km, and irradiated with a neutrino flux emanating from the surface of the neutrino-sphere into the shock tube. Interaction due to electron-neutrino scattering off of fluid nucleons is used to account for neutrino heating of the fluid.}
   {The numerical method used is the Godunov method implementing an exact Riemann solver \citep*{rezzolla} when needed. The simulated core-collapse produces an energy output of up to $10^{36}$~J on a timescale of $3.33$~s. It was found that the Godunov method performs very well using \citep*{sod} data, with standard deviations of between $0.22-0.04$~\%. Thus, there is high confidence that the exact Riemann solver is working correctly and that the Godunov method is also reproducing the exact solutions to the Sod data with high confidence.}
   {A runtime of $3.33$~s was used. The objective was to determine what would happen to the shock system over longer timescales, but the simulations showed an explosion in a much shorter time frame. It was found that the shock not only continues to move outward, but is also driven by large energy and pressure behind the shock. The density profiles show that fluid mass is pushed outward, but the fluid velocity actually reversed on timescales close to $3.33$~s. In this paper, net mass movement occurs at $500-600$~km and on a longer timescale of $3.33$~s mass has moved out to $700$~km. This supports the \emph{delayed post bounce shock mechanism}.}
   {}
\end{abstract}

\keywords{core-collapse supernova --
          neutrino heating in supernova --
          delayed post bounce shock mechanism --
					prompt post bounce shock mechanism
         }
%

\section{Introduction}

The discovery of weak neutral currents and the corresponding importance of neutrino scattering off nucleons lead to the realization that the forming neutron star is highly opaque to neutrinos. Thus, the neutrino luminosities were too low and the energy transfer rate was not large enough to invert the infall of the surrounding gas into an explosion \citep*{janka}. One approach to understanding the role of the neutrinos and the particulars of the mechanisms of the energy transfers is to consider neutrino cross-sections for interaction with the stellar fluid. \citep*{tubbs} and \citep*{bowers} are excellent resources for the neutrino cross-sections of the three flavors of neutrino type, and for the development of the particulars of the neutrino interactions.

For a number of years efforts were concentrated on the \emph{prompt post bounce shock mechanism}, which is the process whereby the energy given to the hydrodynamical shockwave in the moment of core bounce was thought to lead directly to the ejection of the stellar mantle and envelope \citep*{janka}. More realistic models showed that, due to severe energy loss experienced by the shock, the shock's outward propagation stops well inside the iron core \citep*{bruenn}.

This paper's model is based on the work done by \citep*{kuroda}, and also used restricted versions of the equations provided by \citep*{matteo}, \citep*{liu} and \citep*{zhang}. The simulation is that of the behavior of stellar fluid undergoing neutrino heating in a shock tube. The model is planar, using the usual special relativistic hydrodynamics equations.

Only electron-neutrinos are considered, since in gravitational collapse problems the electron-neutrino emission outnumber the production of other types by the dominance of $e^{{}-} + e^{{}+} \rightarrow \nu _{e} + \bar{\nu} _{e}$ \citep*{tubbs}. The numerical method used is the Godunov Finite Volume Method, utilizing an exact Riemann solver on the nodes (boundaries of cells). The exact Riemann solver is a relativistic one, written by \citep*{rezzolla}, which was an improvement on work done by \citep*{pons}.

The physical model is that of stellar fluid undergoing neutrino heating within a 1-dimensional shock tube placed against the surface of the neutrino-sphere. The tube extends $1000$~km out from this location. The coordinate system is Cartesian. The isolation of the system from its surroundings make it an excellent test tube. The results of \citep*{kuroda} were used to set up initial conditions for the simulation when the neutrino model was activated and runs done for each of three cases: the fluid under heating provided by only a neutrino energy flux, the fluid under heating provided by only a neutrino momentum flux and the fluid under heating provided by both neutrino energy flux and neutrino momentum flux.

A runtime of $3.33$~s was used. This is well outside the $0.1$~s quoted by \citep*{janka}. The objective was to determine what would happen to the shock system over longer timescales, but the simulations showed an explosion in a much shorter time frame. It was found that the shock not only continues to move outward, but is also driven by large energy and pressure behind the shock. The density profiles show that fluid mass is pushed outward, but the fluid velocity actually reversed on timescales of $3.33$~s. According to \citep*{bruenn}, shocks stall at $100-200$~km. In this paper, such a stall was observed; in fact, net mass movement occurs at $500-600$~km and on a longer timescale of $3.33$~s mass has moved out to $700$~km. This supports the delayed post bounce shock mechanism.

The Godunov method was thoroughly tested against this solver using shocks moving left to right and right to left, and also at velocities close to the speed of light. It was found that the Godunov method performs very well using Sod data, with standard deviations of between $0.22-0.04$~\%. Thus, there is high confidence that the exact Riemann solver is working correctly and that the Godunov method is also reproducing the exact solutions to the Sod data with high confidence.

It is worthwhile to note that this trim model and code has reproduced the findings of much more complicated 3-dimensional codes running much more realistic neutrino equations of state and detailed microphysics. This paper has confirmed that on longer timescales there is no stalling of the shock system. Instead there is an explosion which reach energies in the $10^{36}$~J range.

%

\section{1-D Special Relativistic Neutrino Hydrodynamics}

\citep*{kuroda} begin with the conservation of total energy-momentum, comprised of the fluid and neutrino tensors.

\begin{equation}\label{E:energyMomentumComponents}
	T^{\alpha \beta} = T^{\alpha \beta}_{(fluid)} + T^{\alpha \beta}_{(\nu)}
\end{equation}

\noindent
Conservation of energy-momentum gives:

\begin{equation}\label{E:kktStress}
	\nabla _{\alpha} T^{\alpha \beta}_{(total)} = \nabla _{\alpha} T^{\alpha \beta}_{(fluid)} + \nabla _{\alpha} T^{\alpha \beta}_{(\nu)} = 0
\end{equation}

\noindent
This implied that Equation~(\ref{E:kktStress}) can be written as,

\begin{equation}
	\nabla _{\alpha} T^{\alpha \beta}_{(fluid)} = -Q^{\beta} \label{E:fluidWithQ}
\end{equation}

\begin{equation}
	\nabla _{\alpha} T^{\alpha \beta}_{(\nu)} = Q^{\beta} \label{E:neutrinoWithQ}
\end{equation}

\noindent
where $Q^{\beta}$ represents the source terms describing the exchange of energy and momentum between the fluid and the neutrino radiation.

The contribution of neutrino mass to the fluid is ignored. Similarly the fluid does not create neutrinos and therefore the fluid does not lose mass. These approximations are reasonable in the outer regions of a supernova where neutrinos are free-streaming. 

\section{The Neutrino Model}

The neutrino-sphere is the region where $\tau_{(\nu)} \geq \frac{2}{3}$. Using the graphs in the work done by \citep*{kuroda}, it was determined that the best position to place the Riemann shock tube for the purposes of this paper was on the surface of the neutrino-sphere. This is the only location which, according to \citep*{kuroda} data, would capture a shock in the arbitrarily chosen time of $40$~ms.

\begin{figure}[!ht]
	\centering
		\includegraphics[scale=0.6]{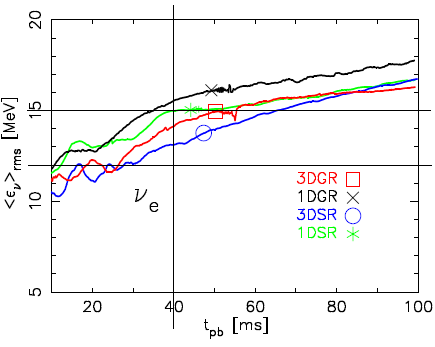}
	\caption[Electron-Neutrino Energy.]{Electron-neutrino energy at $40$~ms.}
	\label{Fig:kktEv}
\end{figure}

\begin{figure}[!ht]
	\centering
		\includegraphics[scale=0.6]{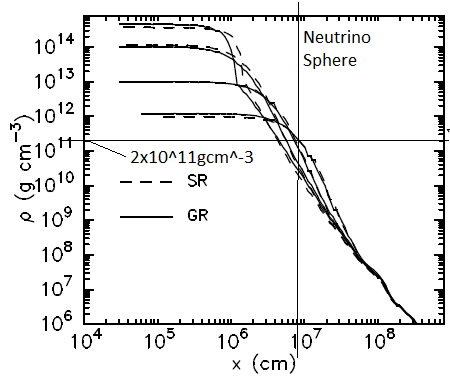}
	\caption[Electron-Neutrino Density.]{Electron-neutrino density at the surface of the neutrino-sphere.}
	\label{Fig:kktRho0}
\end{figure}

\begin{figure}[!ht]
	\centering
		\includegraphics[scale=0.6]{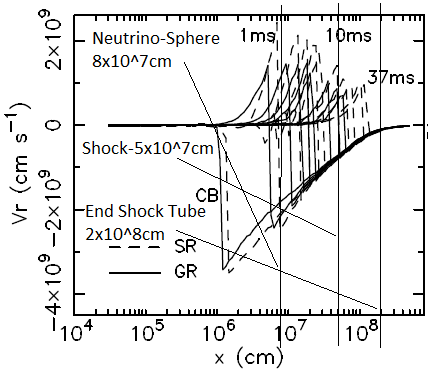}
	\caption[Electron-Neutrino Velocity.]{Electron-neutrino velocity at the surface of the neutrino-sphere.}
	\label{Fig:kktVel}
\end{figure}

The time of $40$~ms was chosen as it was the longest interval for which good data from \citep*{kuroda} graphs could be extracted. The neutrino-sphere's surface corresponds to $x = 800$~km. The length of the shock tube is $1.8 \times 10^{2}$~km. The data of \citep*{kuroda} is used to obtain the values for the velocity profile (Figure~\ref{Fig:kktVel}), the density profile (Figure~\ref{Fig:kktRho0}) and the neutrino specific energy (Figure~\ref{Fig:kktEv}) at time = $10$~ms to time = $40$~ms. The time of $40$~ms was chosen because it intersects the energy profile for a 1-dimensional special relativistic simulation at an energy of $15$~MeV which, according to \citep*{burrows} is in the range of energies for electron-neutrinos in the core collapse. Also, the velocity profile only extends to $37$~ms so no greater time data could be utilized.

Knowing the time constraint and the location of the neutrino-sphere as detailed in \citep*{kuroda} then using the velocity profile allowed the determination of the length of the shock tube and the location of the initial shock (as midway between the neutrino-sphere and the location at a time of $37$~ms).

These considerations can be viewed graphically, as above, by using the work of \citep*{kuroda} and imposing the beginning of the surface of the neutrino-sphere at $10^{5}$~m from the center of the star.

The 1-dimensional versions of \citep*{kuroda} results, with $W = \frac{1}{\sqrt{1 - v^{2}}}$, are;

\begin{equation}
	Q^{t} = e^{- \beta _{\nu} \tau _{\nu}} \left( \epsilon _{s_{\nu}} \right) ^{2} \tilde{\kappa} \left( WE_{(\nu)} - F_{(\nu)} v \right) \label{E:qUpt}
\end{equation}

\begin{equation}
	- Q^{x} = e^{- \beta _{\nu} \tau _{\nu}} \left( \epsilon _{s_{\nu}} \right) ^{2} \tilde{\kappa} \left( - WF_{(\nu)} + P_{(\nu)} v \right) \label{E:qUpx}
\end{equation}

\noindent
where $E_{(\nu)}, F_{(\nu)}$ and $P_{(\nu)}$ are the radiation energy, flux and pressure. Equations~(\ref{E:qUpt}, \ref{E:qUpx}) follow directly from the work of \citep*{kuroda}, where they split the source term, $Q^{\mu}$ as:

\begin{equation}\label{E:kurodaSourceTerm}
 Q^{\mu} \equiv Q^{\mu, C} - Q^{\mu, H} ,
\end{equation}

\noindent
where ``C'' implies a cooling term and ``H'' implies a heating term. This paper is only concerned with the heating term, and neglects any cooling which may occur. This is a constraint on this paper's model.

The heating term is written, by \citep*{kuroda}, as:

\begin{equation}\label{E:kurodaHeatingTerm}
 Q^{\mu, H} = \sum_{\nu \in \nu_{e}, \overline{\nu}_{e}} e^{- \beta _{\nu} \tau _{\nu}} \left( \epsilon _{s_{\nu}} \right) ^{2} \tilde{\kappa}_{\nu} \left( -\mathcal{J}_{\nu} u^{\mu} - \mathcal{H}_{\nu}^{\mu} \right) .
\end{equation}

\noindent
and, also from \citep*{kuroda}, where $\mathcal{J, H}$ denote the Eddington moments in the comoving frame:

\begin{equation}
 \mathcal{J} = E_{(\nu)} W^{2} - 2WF_{(\nu)}^{i}u_{i} + P_{(\nu)}^{ij}u_{i}u_{j} ,
\end{equation}

\begin{equation}
 \mathcal{H}^{\alpha} = \left( E_{(\nu)} W - F_{(\nu)}^{i}u_{i} \right) \left( n^{\alpha} - Wu^{\alpha} \right) + Wh_{\beta}^{\alpha} F_{(\nu)}^{\beta} - h_{i}^{\alpha}u_{j} P_{(\nu)}^{ij} ,
\end{equation}

\noindent
where the Lorentz factor is $W = \alpha u^{0}$ and $h_{\alpha \beta} \equiv g_{\alpha \beta} + u_{\alpha} u_{\beta}$.

Here, $e^{- \beta _{\nu} \tau _{\nu}}$ is taken as $1$, since there is no need to progress smoothly from one region to the next. This paper deals only with the  ``atmosphere'' outside the neutrino-sphere, and simplifies that atmosphere to a region which is optically thin to electron-neutrinos. This means that the neutrino stream is decoupled from the stellar fluid, and so does not adopt the fluid temperature. The neutrino cross section $\sigma _{0}$ means that the neutrino stream presents some opacity to the fluid, and so there can be some neutrino heating.

$\kappa _{sc}$ is defined to be $\kappa _{sc} \equiv \left( \epsilon _{s_{\nu}} \right) ^{2} \tilde{\kappa}$ which simplifies the equations needed to be evolved in order to compute $E_{(\nu)}, F_{(\nu)}$ and therefore $P_{(\nu)}$. So,

\begin{equation}
	Q^{t} = \kappa _{sc} \left( WE_{(\nu)} - F_{(\nu)} v \right) \label{E:qUptAlmostFinal}
\end{equation}

\begin{equation}
	- Q^{x} = \kappa _{sc} \left( - WF_{(\nu)} + P_{(\nu)} v \right) \label{E:qUpAlmostFinal}
\end{equation}


\noindent
$\kappa _{sc} = 1.69 \times 10^{4} \rho _{0}$. Thus, using \citep*{janka} equations and relevant quantity values,

\begin{equation}
	Q^{t} = 1.69 \times 10^{4} \rho _{0} \left( WE_{(\nu)} - F_{(\nu)} v \right) \label{E:qUptFinal}
\end{equation}	

\begin{equation}
	- Q^{x} = 1.69 \times 10^{4} \rho _{0} \left( - WF_{(\nu)} + P_{(\nu)} v \right) \label{E:qUpFinal}
\end{equation}

To implement the evolution of $E_{(\nu)}, F_{(\nu)}$ and $P_{(\nu)}$ is not trivial, and would involve just as much work as the implementation of the fluid evolution. For the purposes of this paper, a simpler model has been developed using the works of \citep*{matteo}, \citep*{liu} and \citep*{zhang}. These works are concerned with gamma-ray bursts, which utilize the same neutrino physics in core-collapse supernova. As such, these works have been adapted to produce the neutrino model for the neutrino heating in this paper.

The primary goal of the neutrino model is an estimation of the neutrino flux. This paper uses the approaches developed by \citep*{matteo}, \citep*{liu} and \citep*{zhang}. The neutrino pressure is given by,

\begin{equation}\label{E:neutrinoPressure}
	P_{(\nu)} = \frac{u_{(\nu)}}{3}
\end{equation}

\noindent
The neutrino energy density, $E_{(\nu)} = u_{(\nu)}$, is given by \citep*{matteo}:

\begin{equation}\label{E:unu}
	E_{(\nu)} = \frac{ \frac{7}{8aT_{(\nu)}^{4}} \left[ \frac{\tau _{(\nu)}}{2} + \frac{1}{\sqrt{3}} \right] } { \frac{\tau _{(\nu)}}{2} + \frac{1}{\sqrt{3}} + \frac{1}{3 \tau _{(\nu)}} }
\end{equation}

\noindent
With the optical depth defined to be $\tau _{(\nu)} = \frac{2}{3}$ at the surface of the neutrino-sphere:

\begin{equation}\label{E:neutrinoEnergyEstimate}
	E _{(\nu)} = \frac{ aT_{(\nu)}^{4} \left[ 7 + 7 \sqrt{3} \right] } { 20 + 8 \sqrt{3} }
\end{equation}

\noindent
where $a$ is the radiation constant and $\sigma$ is the Stefan-Boltzmann constant. The neutrino flux, $F_{(\nu)}$ is given by \citep*{matteo}:

\begin{equation}\label{E:neutrinoFlux1}
	F_{(\nu)} = \frac{\frac{7}{8} \sigma T_{(\nu)}^{4}}{\frac{3}{4} \left( \frac{\tau _{(\nu)}}{2} + \frac{1}{\sqrt{3}} + \frac{1}{3 \tau _{(\nu)}} \right)}
\end{equation}

\noindent
We evaluate the neutrino flux at $\tau _{(\nu)} = \frac{2}{3}$ since this is where the neutrinos are decoupled from the stellar fluid:

\begin{equation}\label{E:neutrinoFluxEstimate}
	F_{(\nu)} = \frac{ 7 \sigma T_{(\nu)}^{4} }{ 5 + 2 \sqrt{3} }
\end{equation}

\noindent
The temperature of the neutrinos, $T_{(\nu)}$, is defined by,

\begin{equation}\label{E:eos1}
	T_{(\nu)} = \varepsilon _{(\nu)} / k_{B}
\end{equation}

\noindent
since the neutrinos are decoupled (this paper assumes that there is total decoupling immediately at $\tau _{\nu} < \frac{2}{3}$) from the stellar fluid outside the neutrino-sphere the temperature is constant. Using the values found earlier, then $T_{(\nu)} = 1.74 \times 10^{11}$~K.

The relevant equations which need to be evolved and used in the calculations at each time step are summarized below. The fluid equations are given with the associated source terms.

\begin{equation}
	D_{, t} + ( Dv^{x} ) _{, x} = 0 \label{E:massConservation1}
\end{equation}

\begin{equation}
	E_{, t} + \left( (E + p) v^{x} \right) _{, x} = - Q^{t}  \label{E:energyWithSource}
\end{equation}

\begin{equation}
	S^{x}_{, t} + \left( S^{x} v^{x} \right) _{, x} = - Q^{x}  \label{E:momentumWithSource}
\end{equation}

\noindent
The source terms are, in geometrized units, and $W = \frac{1}{\sqrt{1 - v^{2}}}$,

\begin{equation}
	Q^{t} = 1.69 \times 10^{4} \rho _{0} \left( WE_{(\nu)} - F_{(\nu)} v \right) \label{E:qUptFinalSummary}
\end{equation}

\begin{equation}
	- Q^{x} = 1.69 \times 10^{4} \rho _{0} \left( - WF_{(\nu)} + P_{(\nu)} v \right) \label{E:qUpFinalSummary}
\end{equation}

\noindent
The quantities $P_{(\nu)}, E_{(\nu)}$ and $F_{(\nu)}$ are given by the equations,

\begin{equation}
	E _{(\nu)} = \frac{ aT_{(\nu)}^{4} \left[ 7 + 7 \sqrt{3} \right] } { 20 + 8 \sqrt{3} } \label{E:neutrinoEnergyDensity}
\end{equation}

\begin{equation}
	P_{(\nu)} = \frac{E_{(\nu)}}{3} \label{E:neutrinoPressure1}
\end{equation}

\begin{equation}
	F_{(\nu)} = \frac{ 7 \sigma T_{(\nu)}^{4} }{ 5 + 2 \sqrt{3} } \label{E:neutrinoFlux}
\end{equation}

\begin{equation}
	T_{(\nu)} = \frac{\varepsilon _{(\nu)}}{k_{B}}
\end{equation}

\noindent
This is the complete set of fluid and neutrino equations implemented in the code in order to simulate a 1-D special relativistic core collapse supernova.

\section{Simulation Results and Analysis}


The work here presented approached the model in three steps. First, only the neutrino energy flux was applied, then only the neutrino momentum flux and finally both fluxes as would be the real case. With only the neutrino energy flux applied, more mass was moved out to the surface of the collapsing star, indicating an impending explosion with some mass imploding onto the neutrino-sphere. When only the neutrino momentum flux was applied, it was observed that mass was driven both outward and inward with a large rarefaction between the two. This indicated a large amount of interaction between the electron-neutrinos and the stellar material.

\begin{figure}[!ht]
	\centering
		\includegraphics[scale=0.5]{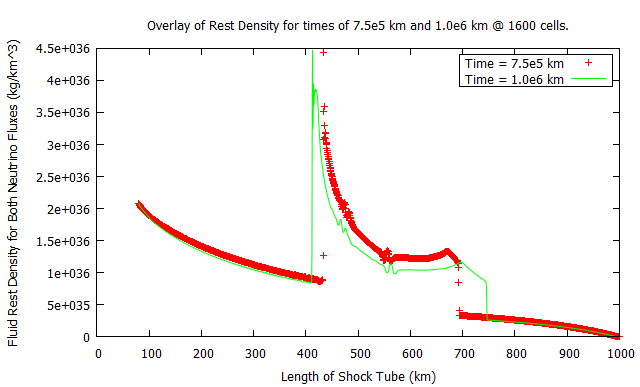}
	\caption[Rest density data.]{Rest density data with both neutrino fluxes on at times of $2.5$~s \& $3.33$~s.}
	\label{Fig:kktTimeOverlayNuOndensityboth}
\end{figure}

\begin{figure}[!ht]
	\centering
		\includegraphics[scale=0.5]{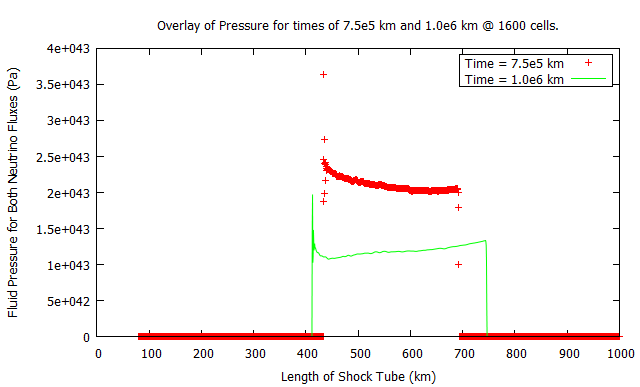}
	\caption[Pressure data.]{Pressure data with both neutrino fluxes on at times of $2.5$~s \& $3.33$~s.}
	\label{Fig:kktTimeOverlayNuOnpressureboth}
\end{figure}

\begin{figure}[!ht]
	\centering
		\includegraphics[scale=0.5]{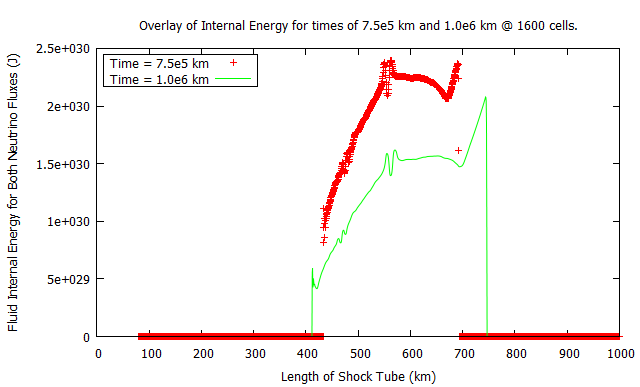}
	\caption[Internal energy data.]{Internal energy data with both neutrino fluxes on at times of $2.5$~s \& $3.33$~s.}
	\label{Fig:kktTimeOverlayNuOnenergyboth}
\end{figure}

When both neutrino fluxes were activated, a reversal of the fluid pressure occurred at the left edge of the pressure profile, while at the right edge the pressure is still very high. This indicated a settling of material onto the neutrino-sphere and an explosive effect at the outgoing wave to the right. So there is both an implosion and an explosion. This is shown in (Figures~\ref{Fig:kktTimeOverlayNuOndensityboth}, \ref{Fig:kktTimeOverlayNuOnpressureboth} \& \ref{Fig:kktTimeOverlayNuOnenergyboth}).

When both fluxes were used, the results showed that the effects from the neutrino momentum flux summed with the effects of neutrino energy flux, resulting in greater fluid pressure and density. This is expected because the source term in the energy-momentum tensor for the neutrinos was split as the sum of the source provided by the neutrino momentum and the source provided by the neutrino energy.

The result of a mixture of implosion and explosion when both the neutrino fluxes were applied and the simulation run for $3.33$~s is intuitively expected. In this scenario, the fluid pressure reverses but the outer shock is still accelerating outward. The inner shock moves inward very little, and a pressure spike is observed. It is conceivable that neutrinos within the neutrino-sphere can be reheated and another burst would be produced into the shock tube, thereby bolstering the outgoing shock and powering the supernova explosion.

\begin{figure}[!ht]
	\centering
		\includegraphics[scale=0.5]{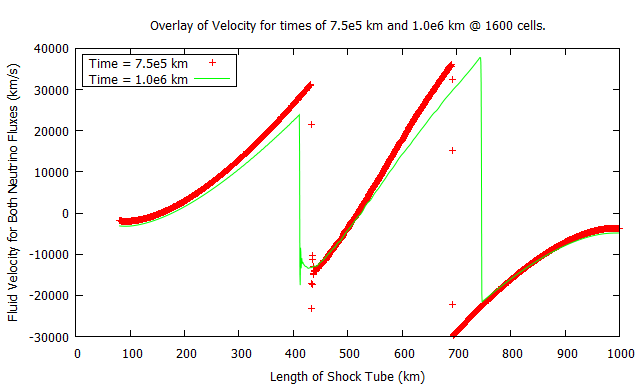}
	\caption[Velocity data.]{Velocity data with both neutrino fluxes on at times of $2.5$~s \& $3.33$~s.}
	\label{Fig:kktTimeOverlayNuOnvelocityboth}
\end{figure}


\section{Conclusions}

This paper supports the \emph{delayed post bounce shock mechanism} and employs a number of ad-hoc terms. The model is a toy model, being in 1-dimension and not taking into account a $\frac{1}{R^{2}}$ term even though the shock tube is very long, extending from the surface of the neutrino-sphere at 8000km to a region in the ``atmosphere'' at 100000km. The other assumption is that outside of the neutrino-sphere, it can be assumed that the optical depth drops to $0$ and remains at $0$ out to the actual atmosphere of the dying star.

This eliminates a e$^{-\tau}$ (where $\tau$ is the optical depth) term in the source term, where the source is the neutrino stream produced from a hot spot in the left side of the shock tube. That source term involves two assumptions. One is that the neutrino momentum flux produced at the hot spot is that value throughout the length of the shock tube, that is, a constant. The second is that the neutrino energy flux is also a constant along the shock tube, due to decoupling of the neutrino stream from the stellar fluid outside of the neutrino-sphere. This made that source term very easy to implement. These assumptions were done in special relativity.

The optical depth needs to be considered carefully, with a full integral of d$\tau = \kappa \rho_{0}$~dx carried out to find $\tau$ along the shock tube. This can then be used to determine e$^{-\tau}$ which would scale the neutrino energy and neutrino flux properly along the shock tube. Finally, the tube itself needs to be changed to a frustum, to better model the spherical symmetry in a 1-dimensional model, which would introduce a $\frac{1}{R^{2}}$ term.

The neutrino momentum and energy fluxes need to be calculated at each time slice using \citep*{shibata} evolution equations for each of these fluxes. This would accurately model the neutrino fluxes along the shock tube, instead of assuming them to be constant once produced at the neutrino-sphere hot spot. The work of \citep*{kuroda} implement these equations. It would be interesting to determine how closely this thesis' results agree with evolved neutrino fluxes. One effect that is known to be resolved in this case is that of the spike at the inner shock. This occurs because of the inclusion in the fluid evolution of a constant neutrino flux. The neutrino evolution equations will correctly control this value, and lead to more stable results.

Realistic neutrino and fluid equations of state need to be used. This thesis used the ideal gas equation of state, both for the fluid and neutrino gas. The work of \citep*{janka} makes it clear that more realistic equations of state produce better neutrino evolution results. This may lead to higher explosion energies; it is not clear if this may actually be the case, but integrating the more realistic equations of state into this paper's code will be a good test ground.

\end{document}